
\input harvmac
\input tables 
\input epsf.tex

\noblackbox

\newcount\figno
\figno=0
\def\fig#1#2#3{
\par\begingroup\parindent=0pt\leftskip=1cm\rightskip=1cm\parindent=0pt
\baselineskip=11pt
\global\advance\figno by 1
\midinsert
\epsfxsize=#3
\centerline{\epsfbox{#2}}
\vskip 12pt
{\it Figure \the\figno:} #1\par
\endinsert\endgroup\par
}
\def\figlabel#1{\xdef#1{\the\figno}}


\def\fund{  \> {\vcenter  {\vbox  
              {\hrule height.6pt
               \hbox {\vrule width.6pt  height5pt  
                      \kern5pt 
                      \vrule width.6pt  height5pt }
               \hrule height.6pt}
                         }
                   }
           \>\> }

\def\antifund{  \> \overline{ {\vcenter  {\vbox  
              {\hrule height.6pt
               \hbox {\vrule width.6pt  height5pt  
                      \kern5pt 
                      \vrule width.6pt  height5pt }
               \hrule height.6pt}
                         }
                   } }
           \>\> }

\def\sym{  \> {\vcenter  {\vbox  
              {\hrule height.6pt
               \hbox {\vrule width.6pt  height5pt  
                      \kern5pt 
                      \vrule width.6pt  height5pt 
                      \kern5pt
                      \vrule width.6pt height5pt}
               \hrule height.6pt}
                         }
              }
           \>\> }



\lref\KOTY{
  T.~Kawano, Y.~Ookouchi, Y.~Tachikawa and F.~Yagi,
  ``Pouliot Type Duality via a-Maximization,''
  Nucl.\ Phys.\  B {\bf 735}, 1 (2006),
  [arXiv:hep-th/0509230].
}

\lref\RD{
  M.~Flato and C.~Fronsdal,
  ``Representations of Conformal Supersymmetry,''
  Lett.\ Math.\ Phys.\  {\bf 8}, 159 (1984).
}

\lref\Mack{
  G.~Mack,
  ``All Unitary Ray Representations of the Conformal Group $SU(2,2)$ with 
  Positive Energy,''
  Commun.\ Math.\ Phys.\  {\bf 55}, 1 (1977).
}

\lref\EM{
  N.~Seiberg,
  ``Electric-Magnetic Duality in Supersymmetric NonAbelian Gauge Theories,''
  Nucl.\ Phys.\ B {\bf 435}, 129 (1995),
  [arXiv:hep-th/9411149].
}

\lref\bound{
  D.~Kutasov, A.~Parnachev and D.~A.~Sahakyan,
  ``Central Charges and $U(1)_R$ Symmetries in ${\cal N} = 1$
  Super Yang-Mills,''
  JHEP {\bf 0311}, 013 (2003)
  [arXiv:hep-th/0308071].
}

\lref\unitarity{
  D.~Kutasov and A.~Schwimmer,
  ``On Duality in Supersymmetric Yang-Mills Theory,''
  Phys.\ Lett.\ B {\bf 354}, 315 (1995)
  [arXiv:hep-th/9505004].
}

\lref\amax{
  K.~Intriligator and B.~Wecht,
  ``The Exact Superconformal $R$-Symmetry Maximizes $a$,''
  Nucl.\ Phys.\ B {\bf 667}, 183 (2003), 
  [arXiv:hep-th/0304128].
}

\lref\AnselmiI{
  D.~Anselmi, J.~Erlich, D.~Z.~Freedman and A.~A.~Johansen,
  ``Positivity Constraints on Anomalies in Supersymmetric Gauge Theories,''
  Phys.\ Rev.\ D {\bf 57}, 7570 (1998)
  [arXiv:hep-th/9711035].
}

\lref\AnselmiII{
  D.~Anselmi, D.~Z.~Freedman, M.~T.~Grisaru and A.~A.~Johansen,
  ``Nonperturbative Formulas for Central Functions of Supersymmetric Gauge
  Theories,''
  Nucl.\ Phys.\ B {\bf 526}, 543 (1998)
  [arXiv:hep-th/9708042].
}

\lref\tHooft{
  G.~'t Hooft,
  ``Naturalness, Chiral Symmetry, and Spontaneous Chiral Symmetry Breaking,''
    in {\it Recent Developments in Gauge Theories}, eds. 't~Hooft 
    {\it et.$\,$al.} (Plenum Press, New York, 1980), 135. 
}

\lref\PSX{
  P.~Pouliot and M.~J.~Strassler,
  ``Duality and Dynamical Supersymmetry Breaking in $Spin(10)$ with a Spinor,''
  Phys.\ Lett.\ B {\bf 375}, 175 (1996), 
  [arXiv:hep-th/9602031].
}

\lref\kawano{
  T.~Kawano,
  ``Duality of ${\cal N}=1$ Supersymmetric $SO(10)$ Gauge Theory with Matter 
    in the Spinorial Representation,''
  Prog.\ Theor.\ Phys.\  {\bf 95}, 963 (1996), 
  [arXiv:hep-th/9602035].
}

\lref\SpinX{
M.~Berkooz, P.~L.~Cho, P.~Kraus and M.~J.~Strassler,
  ``Dual Descriptions of $SO(10)$ SUSY Gauge Theories with Arbitrary Numbers of
  Spinors and Vectors,''
  Phys.\ Rev.\ D {\bf 56}, 7166 (1997),
  [arXiv:hep-th/9705003].
}

\lref\PSVIII{
  P.~Pouliot and M.~J.~Strassler,
  ``A Chiral $SU(N)$ Gauge Theory and its Non-Chiral $Spin(8)$ Dual,''
  Phys.\ Lett.\ B {\bf 370}, 76 (1996), 
  [arXiv:hep-th/9510228].
}

\lref\pouliot{
  P.~Pouliot,
  ``Chiral Duals of Nonchiral SUSY Gauge Theories,''
  Phys.\ Lett.\ B {\bf 359}, 108 (1995),
  [arXiv:hep-th/9507018].
}


\Title{                                \vbox{\hbox{UT-07-16}} }
{\vbox{\centerline{
          Supersymmetric ${\cal N}=1$ Spin(10) Gauge Theory
}\vskip3mm
\centerline{
                with Two Spinors via $a$-Maximization
}}}

\vskip .2in

\centerline{
                 Teruhiko~Kawano and Futoshi~Yagi
}

\vskip .2in 

\centerline{\sl
               Department of Physics, University of Tokyo,
}
\centerline{\sl
                     Hongo, Tokyo 113-0033, Japan
}
\bigskip

\vskip 2cm
\noindent

We give a detailed analysis of the superconformal fixed points 
of four-dimensional ${\cal N}=1$ supersymmetric $Spin(10)$ gauge theory
with two spinors and vectors by using $a$-maximization procedure.

\bigskip\bigskip
\Date{May, 2007}


\newsec{Introduction}

In the previous paper \KOTY, we studied 
four-dimensional ${\cal N}=1$ supersymmetric $Spin(10)$ gauge theory 
with a single chiral superfield $\Psi$ in the spinor representation 
and $N_Q$ chiral superfields $Q^i$ $(i=1,\cdots ,N_Q)$ 
in the vector representation and with no superpotential
at the superconformal infrared (IR) fixed point.
This theory is believed to have a non-trivial IR fixed point 
for $7 \le N_Q \le 21$, 
where the dual description is available \refs{\PSX,\kawano}.

At the IR fixed points, 
since the conformal dimension $D({\cal O})$ of a gauge invariant 
chiral primary operator ${\cal O}$ can be determined by the 
superconformal $U(1)_R$ charge $R({\cal O})$ \RD\ as
\eqn\RtoD{
D({\cal O}) = {3\over2} R({\cal O}),
}
the $U(1)_R$ symmetry in the superconformal algebra plays an important role.

The unitarity requires the conformal dimension 
of a gauge invariant Lorentz scalar to satisfy
$
D({\cal O}) \geq 1,
$
where the equality is satisfied if and only if ${\cal O}$ is free \Mack.
With \RtoD, 
\eqn\boundR{
R({\cal O}) \geq {2\over3}.
}
However, a gauge invariant chiral primary operator 
sometimes appears to violate the inequality \boundR,
when we na\"\i vely assume that the global symmetry at the IR fixed point
is the same as that in the ultraviolet (UV) region. 
It has been argued in \refs{\EM,\bound,\unitarity}\
that the operator ${\cal O}$ decouples 
from the remaining interacting system to become free at the IR fixed point, 
where a new global $U(1)$ symmetry 
which transforms only ${\cal O}$ is enhanced
and the real $U(1)_R$ charge of ${\cal O}$ becomes 2/3 .

The superconformal $U(1)_R$ symmetry can be expressed as a
linear combination of anomaly-free $U(1)$ symmetries as
\eqn\trialR{
U(1)_R = U(1)_{\lambda} + \sum_i x_i U(1)_i,
}
where global $U(1)$ symmetries under which the gaugino has no charge 
are denoted by $U(1)_i$ $(i=1,2,\cdots)$ 
and an anomaly-free $U(1)$ symmetry which 
transforms the gaugino with charge 1 by $U(1)_{\lambda}$.
In order to determine the superconformal $U(1)_R$ symmetry, 
we have to determine the coefficients $x_i$ in \trialR.
In fact, we may use $a$-maximization \amax\ for this purpose.
Following this method, 
we regard $x_i$ in \trialR\ as variables to be determined
and construct the trial $a$-function
\foot{We omit the overall factor $3/32$ 
of the trial $a$-function in this paper,
which does not affect the calculation of the $U(1)_R$ charges.}
\eqn\triala{
a_0 (x_1,x_2,\cdots) = 3{\rm Tr}R^3 - {\rm Tr}R .
}
Each term in the right hand side of \triala\
represents the 't Hooft anomaly \tHooft,
where the charge $R$ is the $U(1)_R$ charge given in terms of $x_i$
in \trialR, but they are not necessarily the superconformal $U(1)_R$ charges
at the IR fixed point.
If there are no accidental symmetries at the IR fixed point,
the 't Hooft anomalies can be evaluated in the UV 
by using the 't Hooft anomaly matching condition
for asymptotically-free theories.
Then, $a$-maximization tells us that the local maximum 
of the function \triala\ gives $x_i$ 
for the superconformal $U(1)_R$ symmetry in \trialR.
 
However, as mentioned above, 
the function \triala\ does not make sense in the range of $x_i$ where 
gauge invariant chiral primary operators seem to violate 
the unitarity bounds \boundR.
It was proposed in the paper \bound\ that, in the range
 where operators ${\cal O}_i$ seem to violate the unitarity bounds \boundR, 
the trial $a$-function should be modified into
\eqn\KPS{
a(x_1,x_2,\cdots) = a_0 + \sum _i 
\left[ - a_{{\cal O}_i} \left( R({\cal O}_i) \right)
+ a_{{\cal O}_i} \left( 2/3 \right) \right] .
}
The function $a_{{\cal O}_i}$ represents the contribution 
from the operator ${\cal O}_i$ to the trial $a$-function
and can be evaluated as
\eqn\aO{
a_{{\cal O}_i}\left( R({\cal O}_i) \right) = d_{{\cal O}_i} 
\left[ 3\left( R({\cal O}_i) - 1 \right)^3
- \left( R({\cal O}_i)-1 \right) \right] ,
}
where $d_{{\cal O}_i}$ is the number of the components 
of the operator ${\cal O}_i$,
and $R({\cal O}_i)$ is the $U(1)_R$ charge of ${\cal O}_i$,
as given in \trialR.
The term $a_{{\cal O}_i}(2/3)$ 
is obtained by substituting the value $R({\cal O}_i)=2/3$ of free fields
into \aO\ to give $2d_{{\cal O}_i}/9$.
The prescription \KPS\ can be interpreted as subtracting the contribution 
which is evaluated under the assumption 
that the operator ${\cal O}_i$ is interacting
and adding the contribution of the operators as free fields.
Thus, by dividing the range of $x_i$ according to 
which operators hit the unitarity bounds and 
by modifying the trial $a$-function as \KPS\ for each range,
we obtain the trial $a$-function 
in the whole range of the variables $x_i$ \refs{\bound,\KOTY}.
The superconformal $U(1)_R$ symmetry could be identified
by the local maximum of this function.

By using the method discussed above, 
we showed in the previous paper \KOTY\
that the meson operator $M^{ij}=Q^iQ^j$ hits the unitarity bound 
and becomes free for $N_Q=7,8,9$.
We also analyzed the IR fixed point by using the electric-magnetic duality
and found that the decoupling of the meson operator 
can be seen more clearly in the magnetic theory.
In the magnetic theory, 
since the meson operator is described by elementary fields,
we do not need the prescription \KPS.
We thus proved the validity of the prescription \KPS\ in the theory.

The magnetic theory is $SU(N_Q-5)$ gauge theory 
with $N_Q$ antifundamentals $\bar{q}_i$, a single fundamental $q$, 
a symmetric tensor $s$, and singlets $M^{ij}$ and $Y^i$,
and its superpotential is given by
\eqn\onespinorpot{
W_{\rm mag} = M^{ij} \bar{q}_i {s}  \bar{q}_j + Y^i q \bar{q}_i + {\rm det}s.
}
where $M^{ij}$ and $Y^i$ correspond to the gauge invariant operators 
$M^{ij}=Q^iQ^j$ and $Y^i=Q^i \Psi^2$ in the electric theory, respectively.
When $M^{ij}$ hits the unitarity bound, 
it decouples from the interacting system, and thus,
the interaction $M^{ij} \bar{q}_i s \bar{q}_j$ 
in \onespinorpot\ becomes irrelevant at the IR fixed point.
This can be checked by evaluating the $U(1)_R$ charge of this term.
Thus, we may identify the remaining interacting system
with the theory without the term $M^{ij} \bar{q}_i s \bar{q}_j$ 
in the superpotential so that we can construct the trial $a$-function 
of this interacting system together with the free meson 
without the prescription \KPS,
but the resulting function is actually identical to \KPS.

We further discussed that,
since the interaction $M^{ij} \bar{q}_i s \bar{q}_j$ 
in \onespinorpot\ vanishes at the IR fixed point,
we do not have the $F$-term condition for $M^{ij}$, and
new massless degrees of freedom corresponding to $\bar{q}_i s \bar{q}_j$ 
appear there. 
The dual of the magnetic theory without the interaction term 
$M^{ij} \bar{q}_i s \bar{q}_j$ is given by the original electric theory 
but with the superpotential 
\eqn\hitpotential{
W = N_{ij} Q^i Q^j,
}
where $N_{ij}$ are additional singlets and
correspond to $\bar{q}_i s \bar{q}_j$.
We found that the IR fixed point of the original theory is identical 
to that of this theory. 
This renormalization group flow can be seen 
in the original electric theory by
introducing auxiliary fields $M^{ij}$ and 
the Lagrange multipliers $N_{ij}$ to give the superpotential
$$
W=N_{ij}(Q^iQ^j-M^{ij}).
$$
We can see that it is the same theory as the original one by integrating out 
$M^{ij}$ and $N_{ij}$.
The equations of motion give the constraints
\eqn\constMN{
M^{ij}=Q^iQ^j, \quad N_{ij}=0.
}
When $M^{ij}$ hits the unitarity bound, 
the interaction $N_{ij} M^{ij}$ becomes irrelevant,
to give rise to the superpotential \hitpotential\ at the IR fixed point,
where the constraints \constMN\ does not exist.
In this way, we find that $a$-maximization and the electric-magnetic duality
reveal the rich dynamics at the IR fixed point.

In this paper, we extend the analysis
to the theory with two spinors and $N_Q$ vectors
and show that the meson operator
$M^{ij} = Q^iQ^j$ decouples from the interacting system 
to become free for $N_Q=6,7$.

This paper is organized as follows:
In section 2, we briefly review the electric-magnetic duality 
in the theory with two spinors, especially about  
the matching of gauge invariant operators.
In section 3, we study which operators become free 
by using $a$-maximization for both electric and magnetic theory.
Section 4 is devoted to summary and discussion.
In the appendices, we discuss the 
gauge invariant operators in both the electric and the magnetic theory.

\newsec{The Electric-Magnetic Duality}

We study four-dimensional ${\cal N}=1$ supersymmetric $Spin(10)$ gauge theory
 with two chiral superfields $\Psi_I$ $(I=1,2)$ in the spinor representation
 and $N_Q$ chiral superfields $Q^i$ $(i=1,\cdots ,N_Q)$ 
in the vector representation and with no superpotential.
From the 1-loop beta function, we find that it is asymptotically free 
for $N_Q \le 19$.
It is believed that the theory has a non-trivial superconformal IR fixed point
for $6 \le N_Q \le 19$, 
where the magnetic dual description exists \SpinX. 

This theory has the anomaly-free global symmetry 
$SU(N_Q) \times SU(2) \times U(1)_F \times U(1)_{\lambda}$, 
and the fields $Q^i$ and $\Psi_I$ have charges 
$\left( {\bf N_Q},{\bf 1}, -4 , 1 \right)$ and 
$\left( {\bf 1},{\bf 2}, N_Q , -1 \right)$, respectively, under the symmetry.
Here, $U(1)_F$ is a global symmetry under which the gaugino have no charge, 
while $U(1)_{\lambda}$ transforms it with charge 1.
If there are no accidental symmetries at the IR fixed point, 
the $U(1)_R$ symmetry in the superconformal algebra should be given as 
a linear combination of these $U(1)$ symmetries as 
\eqn\defx{
U(1)_R = x U(1)_F + U(1)_{\lambda}
}
with some real number $x$.
Thus, the $U(1)_R$ charge of the matter fields can be expressed as
\eqn\RQRP{
R(Q)=-4x+1, \qquad R(\Psi)=N_Q x -1.
}
We will determine the value of $x$ in the next section 
by using $a$-maximization.

As explained in the introduction, 
in order to construct the trial $a$-function in the whole range of $x$,
we need to know gauge invariant chiral primary operators 
at the IR fixed point.
As discussed in appendix A, the gauge invariant generators 
of the classical chiral ring of this theory are given by
\eqn\electricoperator{
\eqalign{
&M^{ij} = Q^{ai} Q^{aj}, 
\cr
&Y^i_X = \Psi_I^T C (\sigma_2\sigma_X)^{IJ} \Gamma^a \Psi_J Q^{ai}, 
\cr
&C^{i_1 \cdots i_3} = 
\Psi_I^T C (\sigma_2)^{IJ} \Gamma^{a_1 \cdots a_3} \Psi_J 
Q^{a_1i_1} \cdots Q^{a_3i_3}, 
\cr
&B^{i_1 \cdots i_5}_X = 
\Psi_I^T C (\sigma_2\sigma_X)^{IJ} \Gamma^{a_1 \cdots a_5} \Psi_J 
Q^{a_1i_1} \cdots Q^{a_5i_5}, 
\cr
&F^{i_1 \cdots i_7} = 
\Psi_I^T C (\sigma_2)^{IJ} \Gamma^{a_1 \cdots a_7} \Psi_J
Q^{a_1i_1} \cdots Q^{a_7i_7}, 
\cr
&{E_2}^{i_1 \cdots i_9}_X = 
\Psi_I^T C (\sigma_2\sigma_X)^{IJ} \Gamma^{a_1 \cdots a_9} \Psi_J
Q^{a_1i_1} \cdots Q^{a_9i_9}, 
\cr
&G = \Psi^T_I C (\sigma_2\sigma_X)^{IJ} \Gamma^a \Psi_J
\Psi^T_K C (\sigma_2\sigma_X)^{KL} \Gamma^a \Psi_L, 
\cr
&H^{i_1 \cdots i_4} = 
\Psi^T_I C (\sigma_2\sigma_X)^{IJ} \Gamma^{a_1 \cdots a_5} \Psi_J
\Psi^T_K C (\sigma_2\sigma_X)^{KL} \Gamma^{a_1} \Psi_L 
Q^{a_2 i_2} \cdots Q^{a_5 i_5}, 
\cr
&{D_0}^{i_1 \cdots i_6} = 
\varepsilon^{a_1 \cdots a_{10}} Q^{a_1i_1} \cdots Q^{a_6i_6} 
W_{\alpha}{}^{a_7a_8} W^{\alpha a_9a_{10}}, 
\cr
&{D_1}_{\alpha}^{i_1 \cdots i_8} = 
\varepsilon^{a_1 \cdots a_{10}} Q^{a_1i_1} \cdots Q^{a_8i_8} 
W_{\alpha}{}^{a_9a_{10}}, 
\cr
&{D_2}^{i_1 \cdots i_{10}} = 
\varepsilon^{a_1 \cdots a_{10}} Q^{a_1i_1} \cdots Q^{a_{10}i_{10}}, 
\cr
&S = {\rm Tr} \, W^{\alpha} W_{\alpha}.
}}
Here, $a$ and $a_1,a_2,\cdots$ are the indices of the gauge group $Spin(10)$,
and the matrix $C$ is the charge conjugation matrix of it. 
The matrices $\sigma_X$ $(X=1,2,3)$ are the Pauli matrices 
for the flavor of the spinors.
Taking account of the number of the antisymmetrized indices 
of the $SU(N_Q)$ global symmetry,
we see that whether each operator exists depends on $N_Q$. 
For example, the operator ${D_2}^{i_1 \cdots i_{10}}$ exists 
only for $N_Q \ge 10$.

\topinsert
\begintable
\|\| $SU(N_Q-3)$ \| $Sp(1)$ \|\| $SU(N_Q)$ \| 
$SU(2)$ \| $U(1)_F$ \| $U(1)_{\lambda}$ 
\nr
\|\| $a,b,\cdots$ \| $\alpha,\beta,\cdots$ \|\| $i,j,\cdots$ 
\| $I,J,\cdots$ \|\| 
\cr
$\bar{q}_{ai}$ \|\| $\antifund$ \| ${\bf 1}$ \|\| $\antifund$ \| ${\bf 1}$ 
\| $2{N_Q-6\over{N}_Q-3}$ \| ${1\over{N}_Q-3}$ 
\nr
$\bar{q}'{}_a{}^{\alpha I}$ \|\| $\antifund$ \| ${\bf 2}$ \|\| ${\bf 1}$
\| ${\bf 2}$ \| $-{2N_Q\over{N}_Q-3}$ \| ${N_Q-2\over{N}_Q-3}$ 
\nr
$q^a_X$ \|\| $\fund$ \| ${\bf 1}$ \|\| ${\bf 1}$ \| ${\bf 3}$ 
\| $-2N_Q {N_Q-4\over{N}_Q-3}$ \| ${3N_Q-10\over{N}_Q-3}$ 
\nr
$s^{ab}$ \|\| $\sym$ \| ${\bf 1}$ \|\| ${\bf 1}$ \| ${\bf 1}$ 
\| ${4N_Q\over{N}_Q-3}$ \| $-{2\over{N}_Q-3}$ 
\nr
$t^{\alpha I}$ \|\| ${\bf 1}$ \| ${\bf 2}$ \|\| ${\bf 1}$ \| ${\bf 2}$ 
\| $2N_Q$ \| $-2$ 
\nr
$M^{ij}$ \|\| ${\bf 1}$ \| ${\bf 1}$ \|\| $\sym$ \| ${\bf 1}$ \| $-8$ \| $2$ 
\nr
$Y^i_X$ \|\| ${\bf 1}$ \| ${\bf 1}$ \|\| $\fund$ \| ${\bf 3}$ \| 
$2N_Q-4$ \| $-1$
\endtable
\nobreak 
\centerline{\it Table 1: The matter contents of the magnetic theory.}
\endinsert

Now, we turn to the magnetic theory, which is believed to be equivalent 
as the original electric theory in the IR region.
The magnetic theory is given by $SU(N_Q-3) \times Sp(1)$ gauge theory 
with the matter content given by Table 1 and with the superpotential
\eqn\magpot{
W_{\rm mag} = 
M^{ij} \bar{q}_{ai} s^{ab} \bar{q}_{bj} + Y^i_X \bar{q}_{ai} q^a_X 
+ \varepsilon_{\alpha\beta} \varepsilon_{IJ}
\bar{q}'{}_a{}^{\alpha I} s^{ab} \bar{q}'_b{}^{\beta J} 
+  \varepsilon_{\alpha\beta} (\sigma_X \sigma_2)_{IJ}
\bar{q}'{}_a{}^{\alpha I} q^a_X t^{\beta J}.
}
This theory has the same anomaly-free global symmetry as the electric theory.
Thus, the $U(1)_R$ symmetry should also be expressed as 
\defx\ with the same value of $x$ as in the electric theory.

There exist gauge invariant operators in this theory
which correspond to those of the electric theory.
They are fundamental singlets $M^{ij}$ and $Y^i_X$ and 
the following composite operators:
\eqn\matching{
\eqalign{
&(*C)_{i_1 \cdots i_{N_Q-3}} \sim \varepsilon^{a_1 \cdots a_{N_Q-3}}
\bar{q}_{a_1i_1} \cdots \bar{q}_{a_{N_Q-3}i_{N_Q-3}}, 
\cr
&(*B)_{X i_1 \cdots i_{N_Q-5}} \sim
\varepsilon^{a_1 \cdots a_{N_Q-3}} \varepsilon_{\alpha\beta} 
\bar{q}_{a_1i_1} \cdots \bar{q}_{a_{N_Q-5}i_{N_Q-5}}
\bar{q}'{}_{a_{N_Q-4}}^{\alpha I} (\sigma_2 \sigma_X)_{IJ}
\bar{q}'{}_{a_{N_Q-3}}^{\beta J}, 
\cr
&(*F)_{i_1 \cdots i_{N_Q-7}} \sim
\varepsilon^{a_1 \cdots a_{N_Q-3}} 
\varepsilon_{\alpha\beta} \varepsilon_{\gamma\delta} 
\bar{q}_{a_1i_1} \cdots \bar{q}_{a_{N_Q-7}i_{N_Q-7}} 
\cr
&\hskip 3cm \times \bar{q}'{}_{a_{N_Q-6}}^{\alpha I} (\sigma_2 \sigma_X)_{IJ}
\bar{q}'{}_{a_{N_Q-5}}^{\beta J} 
\bar{q}'{}_{a_{N_Q-4}}^{\gamma K} (\sigma_2 \sigma_X)_{KL}
\bar{q}'{}_{a_{N_Q-3}}^{\delta L}, 
\cr
&(*E_2)_{X i_1 \cdots i_{N_Q-9}} \sim
\varepsilon_{a_1 \cdots a_{N_Q-3}} \varepsilon_{XYZ} 
(s\bar{q}_{i_1})^{a_1} \cdots (s\bar{q}_{i_{N_Q-9}})^{a_{N_Q-9}} 
\cr
&\hskip 3cm \times (sw_{\alpha})^{a_{N_Q-8}a_{N_Q-7}} 
(sw_{\alpha})^{a_{N_Q-6}a_{N_Q-5}} q^{N_Q-4}_Y q^{N_Q-3}_Z, 
\cr
&G \sim \varepsilon_{\alpha\beta} t^{\alpha I} (\sigma_2)_{IJ} t^{\beta J}, 
\cr
&(*H)_{i_1 \cdots i_{N_Q-4}} \sim \varepsilon_{IJ}
\varepsilon^{a_1 \cdots a_{N_Q-3}} \varepsilon_{\alpha\beta} 
\bar{q}_{a_1i_1} \cdots \bar{q}_{a_{N_Q-4}i_{N_Q-4}} 
\bar{q}'{}_{a_{N_Q-3}}^{\alpha I} t^{\beta J}, 
\cr
&(*D_0)_{i_1 \cdots i_{N_Q-6}} \sim
\varepsilon_{XYZ} \varepsilon_{a_1 \cdots a_{N_Q-3}} 
(s\bar{q}_{i_1})^{a_1} \cdots (s\bar{q}_{i_{N_Q-6}})^{a_{N_Q-6}}
q^{a_{N_Q-5}}_X q^{a_{N_Q-4}}_Y q^{a_{N_Q-3}}_Z, 
\cr
&(*D_1)_{\alpha i_1 \cdots i_{N_Q-8}} \sim
\varepsilon_{a_1 \cdots a_{N_Q-3}} \varepsilon_{XYZ} 
(s\bar{q}_{i_1})^{a_1} \cdots (s\bar{q}_{i_{N_Q-8}})^{a_{N_Q-8}} 
\cr
&\hskip 3cm  \times (sw^{\alpha})^{a_{N_Q-7}a_{N_Q-6}} 
q_X^{a_{N_Q-5}} q_Y^{a_{N_Q-4}} q_Z^{a_{N_Q-3}}, 
\cr
&(*D_2)_{i_1 \cdots i_{N_Q-10}} \sim
\varepsilon_{a_1 \cdots a_{N_Q-3}} \varepsilon_{XYZ} 
(s\bar{q}_{i_1})^{a_1} \cdots (s\bar{q}_{i_{N_Q-10}})^{a_{N_Q-10}} 
\cr
&\hskip 3cm  \times (sw_{\alpha})^{a_{N_Q-9}a_{N_Q-8}} 
(sw^{\alpha})^{a_{N_Q-7}a_{N_Q-6}} 
q_X^{a_{N_Q-5}} q_Y^{a_{N_Q-4}} q_Z^{a_{N_Q-3}}, 
\cr
&S \sim {\rm Tr} \, w_{\alpha} w^{\alpha} ,\quad 
S' \sim	{\rm Tr} \, \tilde{w}_{\alpha} \tilde{w}^{\alpha}.
}}
Here, $w_{\alpha}$ and $\tilde{w}_{\alpha}$ are the field strength 
of the $SU(N_Q-3)$ and $Sp(1)$ gauge groups, respectively,
\foot{The index $\alpha$ of the field strength 
$w_{\alpha}$ and $\tilde{w}_{\alpha}$ is that of Lorentz spinors,
which would not cause any confusion.}
and the operation $*$ represents the Hodge duality with respect to 
the flavor $SU(N_Q)$ indices.
The magnetic theory has two kinds of glueball superfields corresponding to 
the two gauge group factors.
We can check that every operator has the same charges as that of the 
electric theory.

Furthermore, it seems that more gauge invariant generators exist 
in the magnetic theory than in the electric one. 
They are given by
\eqn\magop{
\eqalign{
&U_0 = {\rm det} s, 
\cr
&{U_1}_{XY} = \varepsilon_{a_1 \cdots a_{N_Q-3}} 
\varepsilon_{b_1 \cdots b_{N_Q-3}} s^{a_1b_1} \cdots s^{a_{N_Q-4}b_{N_Q-4}} 
q^{a_{N_Q-3}}_X q^{b_{N_Q-3}}_Y, 
\cr
&{U_2}_{XY} = \varepsilon_{XX_1X_2} \varepsilon_{YY_1Y_2}
\varepsilon_{a_1 \cdots a_{N_Q-3}} \varepsilon_{b_1 \cdots b_{N_Q-3}} 
\cr
&\hskip	2cm \times s^{a_1b_1} \cdots s^{a_{N_Q-5}b_{N_Q-5}} 
q^{a_{N_Q}-4}_{X_1} q^{a_{N_Q}-3}_{X_2} q^{b_{N_Q}-4}_{Y_1} 
q^{b_{N_Q}-3}_{Y_2}, 
\cr
&U_3 = \varepsilon_{X_1X_2X_3} \varepsilon_{Y_1Y_2Y_3}
\varepsilon_{a_1 \cdots a_{N_Q-3}} \varepsilon_{b_1 \cdots b_{N_Q-3}} 
\cr
&\hskip	2cm	\times 	s^{a_1b_1} \cdots s^{a_{N_Q-6}b_{N_Q-6}}
q^{a_{N_Q}-5}_{X_1} q^{a_{N_Q}-4}_{X_2} q^{a_{N_Q}-3}_{X_3}	
q^{b_{N_Q}-5}_{Y_1} q^{b_{N_Q}-4}_{Y_2} q^{b_{N_Q}-3}_{Y_3}, 
\cr
&(*E_0)_{X i_1 \cdots i_{N_Q-5}} =
\varepsilon_{XYZ} \varepsilon_{a_1 \cdots a_{N_Q-3}} (s\bar{q}_{i_1})^{a_1}
\cdots (s\bar{q}_{i_{N_Q-5}})^{a_{N_Q-5}} q^{a_{N_Q-4}}_Y q^{a_{N_Q-3}}_Z, 
\cr
&(*E_1)_{\alpha X i_1 \cdots i_{N_Q-7}} =
\varepsilon_{a_1 \cdots a_{N_Q-3}} \varepsilon_{XYZ} 
(s\bar{q}_{i_1})^{a_1} \cdots (s\bar{q}_{i_{N_Q-7}})^{a_{N_Q-7}} 
\cr
&\hskip	4cm \times (sw^{\alpha})^{a_{N_Q-6}a_{N_Q-5}} q_Y^{a_{N_Q-4}} 
q_Z^{a_{N_Q-3}}, 
\cr
&(*I_0)_{X i_1 \cdots i_{N_Q-4}} =
\varepsilon_{a_1 \cdots a_{N_Q-3}} (s\bar{q}_{i_1})^{a_1} 
\cdots (s\bar{q}_{i_{N_Q-4}})^{a_{N_Q-4}} q^{a_{N_Q-3}}_X, 
\cr
&(*I_1)_{\alpha X i_1 \cdots i_{N_Q-6}} =
\varepsilon_{a_1 \cdots a_{N_Q-3}} (s\bar{q}_{i_1})^{a_1} \cdots 
(s\bar{q}_{i_{N_Q-6}})^{a_{N_Q-6}} (sw^{\alpha})^{a_{N_Q-5}a_{N_Q-4}} 
q_X^{a_{N_Q-3}}, 
\cr
&(*I_2)_{X i_1 \cdots i_{N_Q-8}} =
\varepsilon_{a_1 \cdots a_{N_Q-3}} (s\bar{q}_{i_1})^{a_1} \cdots 
(s\bar{q}_{i_{N_Q-8}})^{a_{N_Q-8}} 
\cr
&\hskip	4cm \times (sw_{\alpha})^{a_{N_Q-7}a_{N_Q-6}} 
(sw^{\alpha})^{a_{N_Q-5}a_{N_Q-4}} q_X^{a_{N_Q-3}}, 
\cr
&(*J_1)_{\alpha i_1 \cdots i_{N_Q-5}} =
\varepsilon_{a_1 \cdots a_{N_Q-3}} (s\bar{q}_{i_1})^{a_1} \cdots 
(s\bar{q}_{i_{N_Q-5}})^{a_{N_Q-5}} (sw^{\alpha})^{a_{N_Q-4}a_{N_Q-3}}, 
\cr
&(*J_2)_{i_1 \cdots i_{N_Q-7}} =
\varepsilon_{a_1 \cdots a_{N_Q-3}} (s\bar{q}_{i_1})^{a_1} \cdots 
(s\bar{q}_{i_{N_Q-7}})^{a_{N_Q-7}} 
\cr
&\hskip	4cm \times (sw_{\alpha})^{a_{N_Q-6}a_{N_Q-5}} 
(sw^{\alpha})^{a_{N_Q-4}a_{N_Q-3}}.
}}
In spite of our best effort, 
we have not succeeded to show that these operators
are decomposed or vanish in the classical chiral ring,
as discussed in Appendix B.

The discrepancy makes it difficult for us to understand what happens 
at the IR fixed point.
Though these two theories might actually not be equivalent to each other
at the IR fixed point, 
it is not plausible that all the other 
non-trivial checks discussed in \SpinX\ are only accidental.
Thus, in this paper, we assume that the classical chiral ring is deformed 
by the quantum effects and that the quantum chiral rings of both the theories
are identical.
However, it is still unclear what is indeed happening quantum-mechanically
at the IR fixed point.
This issue affects the construction of the trial $a$-function.
Therefore, we will consider both the functions 
in the electric and the magnetic theory and compare the results.
In the next section, we will see that both the functions have 
the identical local maximum.

\newsec{$a$-Maximization}

In this section, we study $Spin(10)$ gauge theory with 
two spinors and $N_Q$ vectors at the superconformal IR fixed point 
both in the electric and the magnetic theory by using $a$-maximization. 
We calculate the local maximum of the trial $a$-function 
defined in the whole range of the parameter $x$
and determine which operators become free
at the IR fixed point.

\subsec{$a$-Maximization in the Electric Theory}

We begin with the electric theory.
As the result depends on $N_Q$, we first analyze the case $N_Q=6$.
Taking account of the number of the antisymmetrized indices 
of the global symmetry $SU(N_Q)$, 
we find that the gauge invariant operators 
in this case are $M$, $Y$, $C$, $B$, $G$, $H$, $D_0$, and $S$
in \electricoperator.
Since the $U(1)_R$ charge of the glueball superfield $S$ is always 2 
and never hit the unitarity bound, 
we can concentrate on the other seven operators.
Using \RQRP, the $U(1)_R$ charges $R({\cal O})$ 
of the gauge invariant operators can be written in terms of $x$ as
\eqn\RO{
\eqalign{
&R(M) = -8x+2, \quad R(Y) = 8x-1, \quad R(C) = 1, \quad R(B) = -8x+3 
\cr
&R(G) = 24x-4, \quad R(H) = 8x , \quad R(D_0) = -24x+8.
}}
By solving $R({\cal O}) < 2/3$ for each operator,
we find that the ranges of $x$ are given in figure 1.
Since the operator $C$ does not hit the unitarity bound for 
all the ranges of $x$, it does not appear in the figure.
\fig{\it The ranges of $x$ where each operator hits
	 the unitarity bound for $N_Q=6$.}{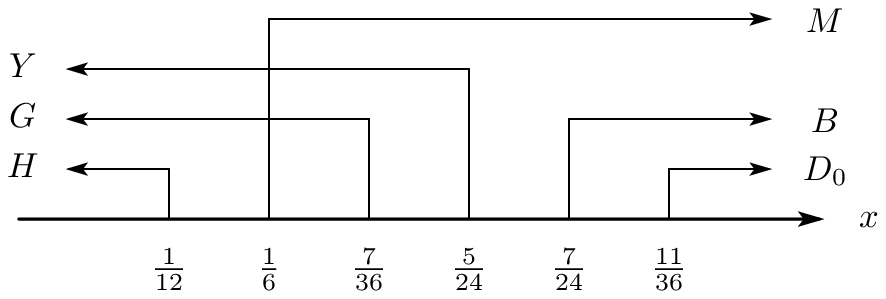}{10 truecm}
\figlabel\ZuI
Now, we construct the trial $a$-function in the whole range of 
the parameter $x$.
The trial $a$-function in the region where
no operators hit the unitarity bound is given by 
$$
a_0(x) = 90 + 32 F[R(\Psi)] + 10 N_Q F[R(Q)],
$$
where $F(y)=3(y-1)^3-(y-1)$.
The first term of this function is the contribution from the gaugino.
The $U(1)_R$ charges $R(\Psi)$ and $R(Q)$ 
may be rewritten in terms of $x$ as \RQRP.
We modify this function as \KPS\ for each range 
according to which operators hit the unitarity bound.
Writing each term in the summation of \KPS\ as 
$f_{\cal O} (x) = - a_{\cal O}\left(R({\cal O})\right) + a_{\cal O}(2/3)$,
the trial $a$-function for the whole range of $x$ is given by
\eqn\trialaele{
a(x) = \left\{
\eqalign{
&a_0(x)+f_{Y}(x)+f_{G}(x)+f_{H}(x),
\hskip 1cm \left( x \le {1\over12} \right) 
\cr
&a_0(x)+f_{Y}(x)+f_{G}(x),
\hskip 1cm \left( {1\over12} \le x \le {1\over6} \right) 
\cr
&a_0(x)+f_{M}(x)+f_{Y}(x)+f_{G}(x),
\hskip 1cm \left( {1\over6} \le x \le {7\over36} \right) 
\cr
&a_0(x)+f_{M}(x)+f_{Y}(x),
\hskip 1cm \left( {7\over36} \le x \le {5\over24} \right) 
\cr
&a_0(x)+f_{M}(x),
\hskip 1cm \left( {5\over24} \le x \le {7\over24} \right) 
\cr
&a_0(x)+f_{M}(x)+f_{B}(x),
\hskip 1cm \left( {7\over24} \le x \le {11\over36} \right) 
\cr	
&a_0(x)+f_{M}(x)+f_{B}(x) + f_{D_0}(R).
\hskip 1cm \left( {11\over36} \le x \right) 
}
\right.
}
More explicitly, the function $f_{\cal O}$ is given by
\eqn\fx{
f_{\cal O} (x) = d_{\cal O} \left\{ - \left[ 3 \left( R({\cal O})-1 \right)^3 
- \left( R({\cal O})-1 \right) \right] + 2/9 \right\},
}
where $d_{\cal O}$ is the number of the components of the operator ${\cal O}$
and is given by
\eqn\dO{
\eqalign{
&d_M = {N_Q(N_Q+1)\over2}, \quad d_Y = 3N_Q , \quad 
d_B = {3 \, N_Q!\over5!(N_Q-5)!} 
\cr
&d_G = 1 , \quad d_H = {N_Q!\over4!(N_Q-4)!} , \quad 
d_{D_0} = {N_Q!\over6!(N_Q-6)!}.
}}
The $U(1)_R$ charge $R({\cal O}_i)$ for each operator ${\cal O}_i$
is given in \RO.
We find that the function \trialaele\ has a unique local maximum at
\eqn\hitx{
x={18N_Q+6-\sqrt{-4N_Q^3+143N_Q^2-928N_Q+1824}\over6(N_Q^2+8N_Q-12)},
}
or equivalently, substituting this to \RQRP,
\eqn\hitR{
\eqalign{
&R(Q) = {3N_Q^2-12N_Q-48+2\sqrt{-4N_Q^3+143N_Q^2-928N_Q+1824}
\over3(N_Q^2+8N_Q-12)}, 
\cr
&R(\Psi) = {12N_Q^2-42N_Q+72-N_Q\sqrt{-4N_Q^3+143N_Q^2-928N_Q+1824}
\over6(N_Q^2+8N_Q-12)},
}}
which is in the range 
where only the meson operator $M^{ij}$ hits the unitarity bound.
Thus, we find that the meson operator $M^{ij}$ decouples 
from the interacting system to become free at the IR fixed point for $N_Q=6$.

Also in the case of $N_Q=7$, we find that $M^{ij}$ hits the unitarity bound
and the $U(1)_R$ charges are given by \hitR\ in the same way as for $N_Q=6$,
though the ranges of $x$ are different from figure \ZuI.

\fig{\it The ranges of $x$ where each operator hits
	 the unitarity bound for $N_Q=8$.}{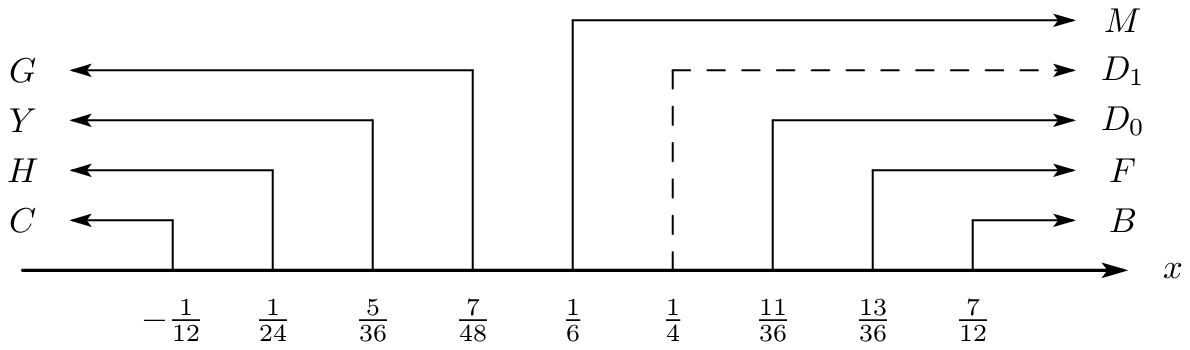}{12 truecm}
\figlabel\ZuII
We go on to the case $N_Q = 8$.
The ranges of $x$ are divided as figure 2. 
In this case, we encounter a subtlety that
we do not understand how to deal with the situation
where gauge invariant Lorentz spinors like $D_{1\alpha}$ 
hit the unitarity bound
\foot{The unitarity bound for gauge invariant Lorentz spinors is 
$R({\cal O}) \ge 1$ \Mack.}.
The best we can do at this stage is 
just to neglect them assuming that such operators are massive in this case
as in the previous paper \KOTY.
Even if they are actually massless, 
our analysis in the region where they do not hit the unitarity bound,
which is $x \le 1/4$ for this case, is still valid. 
We find that the trial $a$-function have a unique local maximum at 
\eqn\nohitx{
x= {12N_Q-\sqrt{2900-N_Q^2}\over6(N_Q^2-20)},
}
or equivalently,
\eqn\nohitR{
\eqalign{
&R(Q) = {3N_Q^2-24N_Q-60+2\sqrt{2900-N_Q^2}\over3(N_Q^2-20)}, 
\cr
&R(\Psi) = {6N_Q^2+120-N_Q\sqrt{2900-N_Q^2}\over6(N_Q^2-20)}.
}}
This is in the range where no operators hit the unitarity bounds.
Though the ranges of $x$ depend on $N_Q$,
we obtain the similar result for $9\le N_Q \le 19$.

In summary, we find that for $N_Q=6,7$, 
the $U(1)_R$ charges are given by \hitR\
and the meson operator $M^{ij}$ becomes free,
while for $8 \le N_Q \le 19$, 
the $U(1)_R$ charges are given by \nohitR\
and no operators become free.

\subsec{$a$-Maximization in the Magnetic Theory}

We next study the magnetic theory. 
Though we expect the same results as that in the electric theory, 
it is non-trivial because of the extra operators \magop.
The trial $a$-function of the magnetic theory is 
different from that of the electric theory 
in the region where such operators hit the unitarity bounds.

We begin with the case $N_Q=6$ and compare with the result 
of the previous subsection.
The gauge invariant operators are
$U_0$, $U_1$, $U_2$, $E_0$, $I_0$, $I_1$, and $J_1$ in \magop,
which exist only in the magnetic theory, as well as 
$M$, $Y$, $C$, $B$, $G$, $H$, $D_0$, and $S$ in \matching,
which have the counterpart in the electric theory.
The charge of these operators can be written with $x$ of \defx\
by using the charges of $U(1)_F$ and $U(1)_{\lambda}$ for each field
given in Table 1.
They are given by \RO\ and also by
\eqn\ROmag{
\eqalign{
&R(U_0) = 24x - 2 , \quad R(U_1) = 4 , \quad R(U_2) = -24x + 10, 
\quad R(E_0) = -8x+5 
\cr
& R(I_0) = 8x+2 , \quad R(I_1) = 3 , \quad R(J_1) = 16x.
}}
We thus, find that the ranges of $x$ where each operator
hits the unitarity bound is given by figure 3.
Since the operators $C$, $U_1$ and $I_1$ do not hit the unitarity bounds for 
all the ranges of $x$, they do not appear in the figure 3.
The bold arrows correspond to the operators 
which exist only in the magnetic theory.
The dotted arrows correspond to the Lorentz spinor operators,
which we ignore as in the previous subsection.

\fig{\it The ranges of $x$ where each operator hits the unitarity bound 
for $N_Q=6$ in the magnetic theory.}{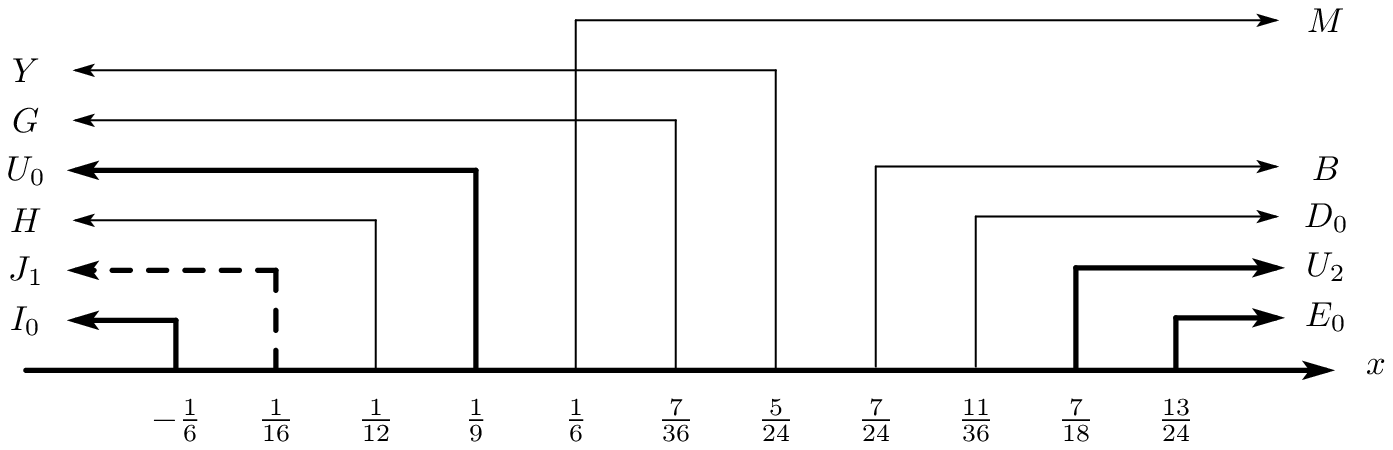}{14 truecm}
\figlabel\ZuIII

As in figure \ZuIII, we find that the trial $a$-function is given by
\eqn\trialamag{
a(x) = \left\{
\eqalign{
&a_0(x)+f_{Y}(x)+f_{G}(x)+f_{U_0}(x)+f_{H}(x)+f_{I_0}(x),
\hskip 1cm \left(  x \le -{1\over6} \right) 
\cr
&a_0(x)+f_{Y}(x)+f_{G}(x)+f_{U_0}(x)+f_{H}(x),
\hskip 1cm \left( -{1\over6} \le x \le {1\over12} \right) 
\cr
&a_0(x)+f_{Y}(x)+f_{G}(x)+f_{U_0}(x),
\hskip 1cm \left( {1\over12} \le x \le {1\over9} \right) 
\cr
&a_0(x)+f_{Y}(x)+f_{G}(x),
\hskip 1cm \left( {1\over9} \le x \le {1\over6} \right) 
\cr
&a_0(x)+f_{M}(x)+f_{Y}(x)+f_{G}(x),
\hskip 1cm \left( {1\over6} \le x \le {7\over36} \right) 
\cr
&a_0(x)+f_{M}(x)+f_{Y}(x),
\hskip 1cm \left( {7\over36} \le x \le {5\over24} \right) 
\cr
&a_0(x)+f_{M}(x),
\hskip 1cm \left( {5\over24} \le x \le {7\over24} \right) 
\cr
&a_0(x)+f_{M}(x)+f_{B}(x),
\hskip 1cm \left( {7\over24} \le x \le {11\over36} \right) 
\cr
&a_0(x)+f_{M}(x)+f_{B}(x) + f_{D_0}(x),
\hskip 1cm \left( {11\over36} \le x \le {7\over18} \right) 
\cr
&a_0(x)+f_{M}(x)+f_{B}(x) + f_{D_0}(x) + f_{U_2}(x),
\hskip 1cm \left( {7\over18} \le x \le {13\over24} \right) 
\cr
&a_0(x)+f_{M}(x)+f_{B}(x) + f_{D_0}(x) + f_{U_2}(x) + f_{E_0}(x),
\hskip 1cm \left( {13\over24} \le x \right) 
}
\right.
}
where $f_{\cal O} (x)$ is given by \fx.
The numbers of the components $d_{\cal O}$ which appear in \fx\
are given by \dO\ and also by
\eqn\dOmag{
d_{U_0} = 1, \quad
d_{U_2} = 6 , \quad
d_{E_0} = {3 \, N_Q!\over5!(N_Q-5)!} , \quad
d_{I_0} = {3 \, N_Q!\over4!(N_Q-4)!}.
}
For the range $1/9 \le x \le 7/18$, 
where the operators which exist only in the magnetic theory 
do not hit the unitarity bound,
the trial $a$-function \trialamag\ have 
the same shape as that of the electric theory.
As the trial $a$-function \trialaele\ of the electric theory 
have a local maximum in this range, this function also have 
the local maximum at the same value of $x$.
We can also check that there are no other local maximum 
throughout the whole range of $x$,
though the function itself is different from that in the electric theory.
Also in the case of $N_Q=7$, we can obtain the same result
as in the electric theory.

\fig{\it The ranges of $x$ where each operator hits the unitarity bound 
for $N_Q=8$ in the magnetic theory.}{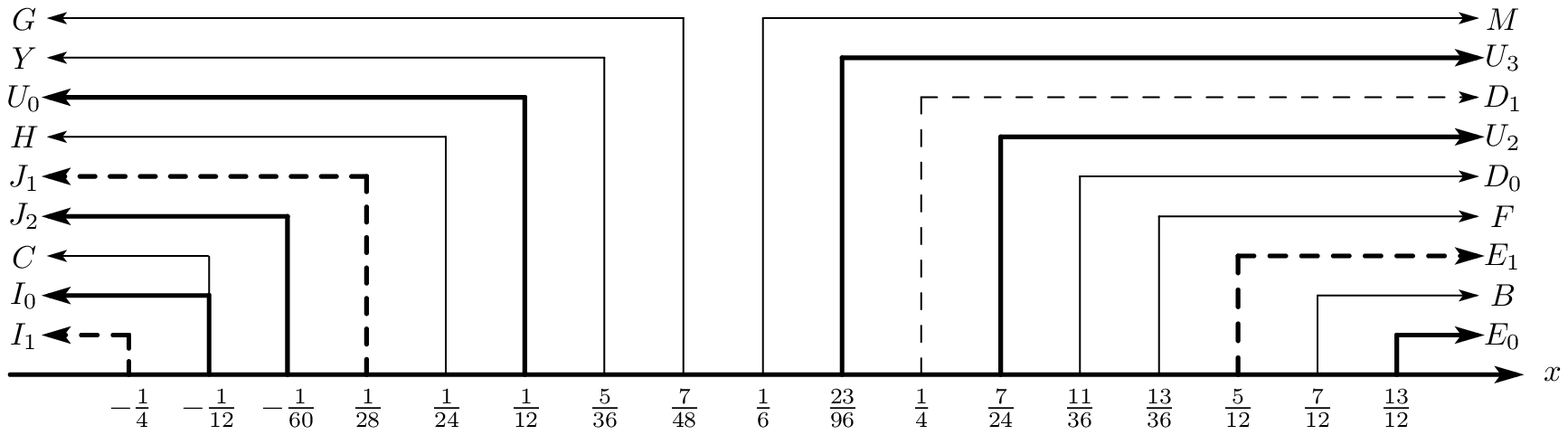}{16 truecm}
\figlabel\ZuIV
In the case of $N_Q=8$, the ranges of $x$ are given in figure 4.
We find that the trial $a$-function has the same shape 
as that of the electric theory for $1/12 \le x \le 23/96$,
which includes the local maximum given by \nohitx.
Since we can verify that there are no local maximum outside this range,
we find that the trial $a$-function have the unique local maximum,
and no operators become free.
Though the ranges of $x$ depend on $N_Q$,
we can find the same result as in the electric theory
also for $9\le N_Q \le 19$.

Thus, we obtain the same results about the value of the $U(1)_R$ charge 
in spite of the discrepancy of the gauge invariant operators.

\newsec{Summary and Discussion}

We have studies $Spin(10)$ gauge theory 
with $N_Q$ vectors and two spinors.
We found that the meson operator $M^{ij}=Q^iQ^j$ decouples from the 
interacting system to become free for $N_Q=6,7$.

We have discussed the renormalization group flow
for the single spinor case in the paper \KOTY.
In particular, for $N_Q=7,8,9$, we have seen the two 
electric theories flow into the same theory at the IR fixed point.
In the present case, since the magnetic theory flows into the theory
without the term $M^{ij} \bar{q}_i s \bar{q}_j$ in the IR,
the electric theory flows into the same theory as that
with $N_{ij} Q^i Q^j$ in the superpotential,
as discussed for the single spinors case in the introduction 
(see \KOTY\ for more details).


\vskip 0.5in
\bigskip

\centerline{\bf Acknowledgements}
We are indebted to Yutaka Ookouchi for collaboration at
the early stages of this work. The research of T.K. was supported in part 
by the Grants-in-Aid (\#16740133) and (\#19540268) from the MEXT of Japan.
The research of F.Y. was supported in part 
by JSPS Research Fellowships for Young Scientists.


\appendix{A}{Gauge Invariant Operator of The Electric Theory}

In this appendix, we explain how to obtain the 
gauge invariant generators \electricoperator\
of the classical chiral ring of the electric theory.
In order to deal with the operators including the $Spin(10)$ spinors $\Psi_I$,
let us recall that the product of the spinors can be decomposed into 
antisymmetric tensor representations as
$$
{\bf 16} \times {\bf 16} = [1] + [3] + [5]_+
$$
where $[n]$ represents the rank $n$ antisymmetric tensor,
and the rank $5$ tensor is self-dual.
They can be explicitly expressed as 
$
\Psi^T_I C \Gamma^{a_1 \cdots a_n} \Psi_J.
$
These are symmetric under the exchange of $I$ and $J$ for $n=1,5$ and 
antisymmetric for $n=3$. 
All gauge invariant operators can be obtained by contracting the 
$Spin(10)$ gauge indices $a_i(=1,\cdots 10)$
of the antisymmetric tensors $\Psi^T_I C \Gamma^{a_1 \cdots a_n} \Psi_J$
$(n=1,3,5)$, the vectors $Q^{ai}$,
the field strength $W_{\alpha}^{a_1a_2}$, 
and the antisymmetric invariant tensors $\varepsilon^{a_1 \cdots a_{10}}$.
However, many of the operators constructed in this way 
are decomposed into the product of other gauge invariant operators
or vanish up to the $\bar{D}^2$ exact term.
In order to identify the independent gauge invariant operators, 
we discuss the constraints among the chiral fields 
$Q$, $\Psi$, $W_{\alpha}$, 
and the invariant tensors $\varepsilon^{a_1 \cdots a_{10}}$.

Since the invariant tensor $\varepsilon^{a_1 \cdots a_{10}}$ satisfies
\foot{The brackets $[ \quad ]$ denote the antisymmetrization of the indices.}
\eqn\ConI{
\varepsilon^{a_1 \cdots a_{10}} \varepsilon_{b_1 \cdots b_{10}} 
= \delta^{a_1}_{[b_1} \cdots \delta^{a_{10}}_{b_{10}]},
}
we can see that a pair of the invariant tensors can annihilate.
Therefore, all the gauge invariant operators can be reduced 
into those with at most one of the invariant tensor
$\varepsilon^{a_1 \cdots a_{10}}$.

It follows from \ConI\ that 
\eqn\ConII{
\varepsilon^{a_1 \cdots a_{10}} \Psi^T_I C \Gamma^{b_1 \cdots b_n} \Psi_J 
\propto \delta^{[a_1}_{b_1} \cdots \delta^{a_n}_{b_n}
\Psi^T_I C \Gamma^{a_{n+1} \cdots a_{10}]} \Psi_J.
}
If we introduce the antisymmetric tensors of rank 7 and 9,
as seen from \ConII, we do not need operators with both of 
the invariant tensor $\varepsilon^{a_1 \cdots a_{10}}$
and the antisymmetric tensor.
Thus, we find that all the invariants are classified 
into operators containing no spinors
with at most one of the invariant tensors 
$\varepsilon^{a_1 \cdots a_{10}}$ 
and those with spinors and none of the invariant tensors 
$\varepsilon^{a_1 \cdots a_{10}}$.

We begin with the operators in the former class.
A constraint
between the field strength $W_{\alpha}$ and other fields $q$ is given by
\eqn\ConIII{
W_{\alpha}^A (T^A)^a{}_b \, q^b 
\propto \bar{D}^2 \left[ e^{-V} D_{\alpha} ( e^V q )\right]^a
\sim 0, 
}
where $q$ is a field in a representation of $Spin(10)$ and 
$T^A$ is the generator in the representation.
For example, when it is the field strength $W_{\alpha}$, we obtain that
$
\{ W_{\alpha}, W_{\beta} \} \sim 0.
$
Thus, operators with more than two of the field strength in this class
vanish by the anticommutativity of them.
Taking account of \ConIII, we find that all
the operators in this class are given by
\eqn\eleopI{
\eqalign{
&M^{ij} = Q^{ai} Q^{aj} , \quad
S = {\rm Tr} \, W^{\alpha} W_{\alpha}, 
\cr
&{D_0}^{i_1 \cdots i_6} = 
\varepsilon^{a_1 \cdots a_{10}} Q^{a_1i_1} \cdots Q^{a_6i_6} 
W_{\alpha}{}^{a_7a_8} W^{\alpha a_9a_{10}}, 
\cr
&{D_1}_{\alpha}^{i_1 \cdots i_8} = 
\varepsilon^{a_1 \cdots a_{10}} Q^{a_1i_1}\cdots Q^{a_8i_8} 
W_{\alpha}{}^{a_9a_{10}}, 
\cr
&{D_2}^{i_1 \cdots i_{10}} = 
\varepsilon^{a_1 \cdots a_{10}} Q^{a_1i_1} \cdots Q^{a_{10}i_{10}}.
}}

We go on to the latter class.
We first consider the operators without the field strength.
Most of the constraints on the spinors can be obtained 
from the Fierz identities.
After repeat use of the Fierz identities and lengthy calculations, 
we find that the product
\eqn\product{
\Psi^T_I C \Gamma^{a_1 \cdots a_i c_1 \cdots c_n} \Psi_J 
\Psi^T_K C \Gamma^{b_1 \cdots b_j c_1 \cdots c_n} \Psi_L
}
can in general be given by a linear combination of the products
\eqn\bibilinear{
\eqalign{
&\Psi^T_I C (\sigma_2\sigma_X)^{IJ} \Gamma^a \Psi_J
\Psi^T_K C (\sigma_2\sigma_X)^{KL} \Gamma^a \Psi_L, 
\cr
&\Psi^T_I C (\sigma_2\sigma_X)^{IJ} \Gamma^{a_1 \cdots a_4 b} \Psi_J
\Psi^T_K C (\sigma_2\sigma_X)^{KL} \Gamma^{b} \Psi_L,
}}
and those where two antisymmetric tensors are not at all 
contracted with each other.
The sum of the ranks of the two antisymmetric tensors in the third 
contribution is always less than that of the original product \product.
By using this fact, it turns out that the third contribution
is decomposed into other invariant operators.
When we use the products of the antisymmetric tensors,
they are thus given by \bibilinear.
Therefore, we can see that all the operators with no field strength 
in this class contain at most two of the antisymmetric tensors.
More explicitly, they are given by
\eqn\eleopIII{
\eqalign{
&Y^i_X = \Psi_I^T C (\sigma_2\sigma_X)^{IJ} \Gamma^a \Psi_J Q^{ai}, 
\cr
&C^{i_1 \cdots i_3} = 
\Psi_I^T C (\sigma_2)^{IJ} \Gamma^{a_1 \cdots a_3} \Psi_J
Q^{a_1i_1} \cdots Q^{a_3i_3}, 
\cr
&B^{i_1 \cdots i_5}_X = 
\Psi_I^T C (\sigma_2\sigma_X)^{IJ} \Gamma^{a_1 \cdots a_5} \Psi_J 
Q^{a_1i_1} \cdots Q^{a_5i_5}, 
\cr
&F^{i_1 \cdots i_7} = 
\Psi_I^T C (\sigma_2)^{IJ} \Gamma^{a_1 \cdots a_7} \Psi_J
Q^{a_1i_1} \cdots Q^{a_7i_7}, 
\cr
&{E_2}^{i_1 \cdots i_9}_X = 
\Psi_I^T C (\sigma_2\sigma_X)^{IJ} \Gamma^{a_1 \cdots a_9} \Psi_J
Q^{a_1i_1} \cdots Q^{a_9i_9}, 
\cr
&G = \Psi^T_I C (\sigma_2\sigma_X)^{IJ} \Gamma^a \Psi_J
\Psi^T_K C (\sigma_2\sigma_X)^{KL} \Gamma^a \Psi_L, 
\cr
&H^{i_1 \cdots i_4} = 
\Psi^T_I C (\sigma_2\sigma_X)^{IJ} \Gamma^{a_1 \cdots a_5} \Psi_J
\Psi^T_K C (\sigma_2\sigma_X)^{KL} \Gamma^{a_1} \Psi_L 
Q^{a_2 i_2} \cdots Q^{a_5 i_5}.
}}

We next consider operators with the spinors and the field strength.
The field strength $W_{\alpha}$ in the operators of this class 
only connect to another one $W_{\beta}$ or the antisymmetric tensors due to 
\ConIII\ and $\{ W_{\alpha}, W_{\beta} \} \sim 0$.
By using the identity 
$$
\Gamma^{a_1 \cdots a_m} \Gamma^{bc} 
= \Gamma^{a_1 \cdots a_m bc} 
+ \delta^{[a_1}_{[b}\Gamma^{a_2 \cdots a_m]}{}_{c]}
- \delta^{[a_1 a_2}_{\,\, b \,\, c}\Gamma^{a_3 \cdots a_m]}.
$$
and the relation \ConIII\ for the spinor representation, we obtain
$$
0 \sim W_{bc} \Psi^T_I C \Gamma^{a_1 \cdots a_m bc} \Psi_J 
+ 2 W^{[a_1}{}_{b} \Psi^T_I C \Gamma^{a_2 \cdots a_m] b} \Psi_J 
- 2 W^{[a_1 a_2} \Psi^T_I C \Gamma^{a_3 \cdots a_m]} \Psi_J.
$$
By decomposing this equation into the symmetric and the antisymmetric part 
under the exchange of the flavor indices $I$ and $J$,
we obtain the equations
\eqn\ConIV{
\eqalign{
& W_{bc} \Psi^T_I C \Gamma^{a_1 \cdots a_m bc} \Psi_J \sim 
2 W^{[a_1 a_2} \Psi^T_I C \Gamma^{a_3 \cdots a_m]} \Psi_J, 
\cr
& W^{[a_1}{}_{b} \Psi^T_I C \Gamma^{a_2 \cdots a_m] b} \Psi_J \sim 0.
}}
We can see from the first equation of \ConIV\ that the rank of the 
antisymmetric tensor connected to the field strength with 
two indices can be reduced by four.
By using the second equation of \ConIV,
we find that the operators including the field strength contracted with
two antisymmetric tensors,
$$
\eqalign{
&\Psi^T_I C \Gamma^{a_1 \cdots a_{m-1} b} \Psi_J W^{bc}
\Psi^T_K C \Gamma^{a_1 \cdots a_{m-1} c} \Psi_L , 
\cr
&\Psi^T_I C \Gamma^{a_1 \cdots a_{m-1} b} \Psi_J W^{bc} W^{cd}
\Psi^T_K C \Gamma^{a_1 \cdots a_{n-1} d} \Psi_L ,
}
$$
can be reorganized into the operator where the 
two antisymmetric tensors are directly contracted.
Similarly to the previous discussion leading to \eleopIII,
such products of the antisymmetric tensors can be rewritten,
and if not vanish, 
the field strength is in turn connected to the antisymmetric tensor
with the two indices or is decomposed with another field strength 
into the glueball $S$.
Thus, we find that operators with the spinors and the field strength 
finally vanish according to \ConIII\
or are decomposed into the product of the glueball superfield $S$
and operators with the spinors.

To summarize, the operators in \eleopI, and \eleopIII\ are
the gauge invariant generators
of the classical chiral ring of the electric theory,
as listed in \electricoperator.

\appendix{B}{Gauge Invariant Operators of The Magnetic Theory}

In this appendix, we only discuss the outline on how to obtain the 
gauge invariant generators of the classical chiral ring of the magnetic theory.
Similarly to the case of the electric theory,
an identity about the antisymmetric invariant tensors 
$\varepsilon^{a_1 \cdots a_{N_Q-3}}$, $\varepsilon_{b_1 \cdots b_{N_Q-3}}$
of $SU(N_Q-3)$ gauge group is given by
\eqn\ee{
\varepsilon^{a_1 \cdots a_{N_Q-3}} 
\varepsilon_{b_1 \cdots b_{N_Q-3}} 
= \delta^{a_1}_{[b_1} \cdots \delta^{a_{N_Q-3}}_{b_{N_Q-3}]}.
}
Thus, all the gauge invariant operators can be classified
into operators with none of the antisymmetric invariant tensors,
those with the invariant tensors with the lower indices,
and those with the invariant tensors with the upper indices.

We first consider the operators without the antisymmetric invariant tensors. 
The equation \ConIII\ is also valid for the field strength 
$w_{\alpha}$ and $\tilde{w}_{\alpha}$ of $SU(N_Q-3)$ and $Sp(1)$,
respectively.
Taking \ConIII\ into account together with the $F$-term conditions,
we can verify that operators without the invariant tensors 
are given by the gauge singlets $M$, $Y$, and the composites
\eqn\magopI{
G \sim \varepsilon_{\alpha\beta} t^{\alpha I} (\sigma_2)_{IJ} t^{\beta J},\quad
S \sim {\rm Tr} \, w_{\alpha} w^{\alpha} ,\quad
S' \sim {\rm Tr} \, \tilde{w}_{\alpha} \tilde{w}^{\alpha}.
}
Here, we also have used 
\eqn\sw{
s^{ab} w_b{}^c \sim - s^{cb} w_b{}^a,
}
which follows from \ConIII\ for the symmetric tensors $s^{ab}$.

We next consider operators including the invariant tensors 
$\varepsilon_{a_1 \cdots a_{N_Q-3}}$.
It turns out that all the operators in this class are given by the contraction
of the invariant tensor $\varepsilon_{a_1 \cdots a_{N_Q-3}}$ 
with the four operators
\eqn\combinationI{
\eqalign{
& \quad q^{a_1}_X , \qquad\quad (s\bar{q})^{a_1}{}_i , 
\qquad\quad (sw_{\alpha})^{a_1a_2}, 
\cr
& \quad (sw^n)^{a_1 b_1} \varepsilon_{b_1 \cdots b_{N_Q-3}} ,\quad (n=0,1,2)
}}
which are supposed so that
the indices $a_1$, $a_2$ of the third in \combinationI\ are contracted with 
those of $\varepsilon_{a_1 \cdots a_{N_Q-3}}$,
while the indices $b_2, \cdots, b_{N_Q-3}$
of the other invariant tensor $\varepsilon_{b_1 \cdots b_{N_Q-3}}$ 
in the fourth are contracted with another of \combinationI.
Taking account of the index $X=1,2,3$ of the field $q_X$ and 
the index $\alpha=1,2$ of the field strength $w_{\alpha}$,
we notice that at most three of the first in \combinationI\ and 
two of the third can be contracted with the same invariant tensor.
Therefore, all the operators with the 
single $\varepsilon_{a_1 \cdots a_{N_Q-3}}$ are given by
\eqn\magopII{
D_n, \quad E_n, \quad I_n, \quad J_m , \quad (n=0,1,2 , \quad m=1,2),
}
in \matching\ and \magop.
Note that the operator
$$
(*J_0)_{i_1 \cdots i_{N_Q-3}} =
\varepsilon_{a_1 \cdots a_{N_Q-3}} (s\bar{q}_{i_1})^{a_1} \cdots 
(s\bar{q}_{i_{N_Q-3}})^{a_{N_Q-3}} 
$$
can be decomposed into the product of the operator $C$ in \matching\
and $U_0$ in \magop.

We turn to the gauge invariant operators 
with more than one $\varepsilon_{a_1 \cdots a_{N_Q-3}}$
and find that all the independent gauge invariant operators are given by
\eqn\magopIII{
\eqalign{
&U_0 = {\rm det} s, 
\cr
&{U_1}_{XY} = \varepsilon_{a_1 \cdots a_{N_Q-3}} 
\varepsilon_{b_1 \cdots b_{N_Q-3}} s^{a_1b_1} \cdots s^{a_{N_Q-4}b_{N_Q-4}} 
q^{a_{N_Q}-3}_X q^{b_{N_Q}-3}_Y, 
\cr
&{U_2}_{XY} = \varepsilon_{XX_1X_2} \varepsilon_{YY_1Y_2}
\varepsilon_{a_1 \cdots a_{N_Q-3}} \varepsilon_{b_1 \cdots b_{N_Q-3}} 
\cr
&\hskip	2cm \times s^{a_1b_1} \cdots s^{a_{N_Q-5}b_{N_Q-5}} 
q^{a_{N_Q}-4}_{X_1} q^{a_{N_Q}-3}_{X_2} q^{b_{N_Q}-4}_{Y_1} 
q^{b_{N_Q}-3}_{Y_2},
\cr
&U_3 = \varepsilon_{X_1X_2X_3} \varepsilon_{Y_1Y_2Y_3}
\varepsilon_{a_1 \cdots a_{N_Q-3}} \varepsilon_{b_1 \cdots b_{N_Q-3}} 
\cr
&\hskip	2cm \times s^{a_1b_1} \cdots s^{a_{N_Q-6}b_{N_Q-6}} 
q^{a_{N_Q}-5}_{X_1} q^{a_{N_Q}-4}_{X_2} q^{a_{N_Q}-3}_{X_3}
q^{b_{N_Q}-5}_{Y_1} q^{b_{N_Q}-4}_{Y_2} q^{b_{N_Q}-3}_{Y_3}.
}}
Let us begin with one invariant tensor $\varepsilon_{a_1 \cdots a_{N_Q-3}}$
and all the symmetric tensor $s^{ab}$ contracted with it,
$$
\varepsilon_{a_1 \cdots a_k a_{k+1} \cdots a_{N_Q-3}} 
s^{a_1b_1} \cdots s^{a_kb_k} 
\equiv T_{a_{k+1} \cdots a_{N_Q-3}}{}^{b_1 \cdots b_k},
$$
in an operator of this class.
The indices $a_{k+1}, \cdots, a_{N_Q-3}$ are supposed to be contracted with
those of the first in \combinationI\ 
or those of the field strength $w_{\alpha}$ in the third.
As the indices $b_1, \cdots , b_k$ in $T$ are antisymmetric,
by using \ee, we can rewrite it as
\eqn\se{
T_{a_{k+1} \cdots a_{N_Q-3}}{}^{b_1 \cdots b_k}
\propto T_{a_{k+1} \cdots a_{N_Q-3}}{}^{d_1 \cdots d_k}
\varepsilon_{d_1 \cdots d_k e_{k+1} \cdots e_{N_Q-3}} 
\varepsilon^{b_1 \cdots b_k e_{k+1} \cdots e_{N_Q-3}}.
}
On the other hand, since we are considering the operators with 
more than one invariant tensors, the operators have another invariant tensor
$\varepsilon_{c_1 \cdots c_{N_Q-3}}$ other than those included in \se.
Then, we apply \ee\ again to 
$\varepsilon^{b_1 \cdots b_k e_{k+1} \cdots e_{N_Q-3}}$ in \se\
and $\varepsilon_{c_1 \cdots c_{N_Q-3}}$.
We thus obtain
\eqn\ees{
T_{a_{k+1} \cdots a_{N_Q-3}}{}^{b_1 \cdots b_k} 
\varepsilon_{c_1 \cdots c_{N_Q-3}} 
\propto T_{a_{k+1} \cdots a_{N_Q-3}}{}^{d_1 \cdots d_k}
\varepsilon_{d_1 \cdots d_k [c_{k+1} \cdots c_{N_Q-3}}
\delta^{b_1 \cdots b_k}_{c_1 \cdots c_k]}.
}
After this procedure, other $s^{ab}$ besides those in \ees\ may connect 
to the original $\varepsilon_{a_1 \cdots a_l a_{l+1} \cdots a_{N_Q-3}}$,
upon the use of \ee.
Then, we can use \ee\ for all the symmetric tensors $s^{ab}$ contracted 
with the tensor $\varepsilon_{a_1 \cdots a_l a_{l+1} \cdots a_{N_Q-3}}$
to annihilate the other 
$\varepsilon_{d_1 \cdots d_k c_{k+1} \cdots c_{N_Q-3}}$ in \ees\
and the appearing invariant tensor of the upper indices.
If the resulting operator does not vanish, we obtain the following form
\eqn\esqe{
\varepsilon_{a_1 \cdots a_l a_{l+1} \cdots a_{N_Q-3}}
s^{a_1b_1} \cdots s^{a_lb_l} q^{a_{l+1}} \cdots q^{a_{N_Q-3}}
\varepsilon_{b_1 \cdots b_l b_{l+1} \cdots b_{N_Q-3}},
}
where the remaining indices $ b_{k+1} \cdots b_{N_Q-3}$ are 
contracted with those in \combinationI.
Again, we apply \ee\ to all symmetric tensors contracted with
$\varepsilon_{b_1 \cdots b_k b_{k+1} \cdots b_{N_Q-3}}$ in \esqe\
to eliminate the original invariant tensor 
$\varepsilon_{a_1 \cdots a_{N_Q-3}}$
and the newly appearing invariant tensor.
We find that all the gauge invariant operators 
with more than one $\varepsilon^{a_1 \cdots a_{N_Q-3}}$
except for \magopIII\ vanish or are decomposed into the gauge invariant
operators.

We next consider operators including the invariant tensors 
$\varepsilon^{a_1 \cdots a_{N_Q-3}}$ with the upper indices.
It turns out that all the operators in this class 
are given by the contraction of the invariant tensor 
$\varepsilon^{a_1 \cdots a_{N_Q-3}}$ with the five operators
\foot{The parentheses $( \quad )$ denote the symmetrization of the indices, 
while $[ \quad ]$ does the antisymmetrization.}
\eqn\combinationII{
\eqalign{
&\quad \bar{q}_{a_1 i}, \qquad\quad \varepsilon_{\alpha\beta} \varepsilon_{IJ}
\bar{q}'{}_{a_1}{}^{\alpha I} t^{\beta J}, 
\qquad\quad \varepsilon_{\alpha\beta} \bar{q}'{}_{a_1}{}^{\alpha (I}
\bar{q}'{}_{b}{}^{|\beta| J} q^{|b| KL)} 
\cr
&\quad \varepsilon_{\alpha\beta} \bar{q}'{}_{a_1}{}^{\alpha (I}
\bar{q}'{}_{b_1}{}^{|\beta| J)} \varepsilon^{b_1 \cdots b_{N_Q-3}},
\qquad\quad \varepsilon_{\alpha\beta} \bar{q}'{}_{a_1}{}^{\alpha (I}
\bar{q}'{}_{a_2}{}^{|\beta| J)}
}}
where $q^{a IJ}$ is related to $q^{a X}$ in Table 1 as
$$
q^{a IJ} \equiv q^a_X (\sigma_X \sigma_2)^{IJ},
$$
and thus, it is symmetric under exchange of the indices $I$ and $J$.
The indices $a_1$ and $a_2$ of the fifth operator in \combinationII\ 
are contracted with those of $\varepsilon^{a_1 \cdots a_{N_Q-3}}$,
while $\varepsilon^{b_1 \cdots b_{N_Q-3}}$ 
in the fourth is contracted with the operators in \combinationII.

Taking account of the indices of the local $Sp(1)$ 
and those of the global $SU(2)$,
we find that at most four $\bar{q}'$
can be contracted with $\varepsilon^{a_1 \cdots a_{N_Q-3}}$.
The numbers of the second, the third, the fourth, and the fifth 
operators in \combinationII\ contracted with
the invariant tensor $\varepsilon^{a_1 \cdots a_{N_Q-3}}$ 
are limited from this fact.
Further, if two $\bar{q}'$ from the second, the third, and the fourth
are contracted with the invariant tensor,
the symmetric part of the global $SU(2)$ indices of them
can be rewritten in terms of the fifth and some other parts.
In fact, when the indices of the global $SU(2)$ of
these two $\bar{q}'$ are symmetric,
the local $Sp(1)$ indices of those $\bar{q}'$ should be antisymmetric.
Then, by using the relation for the invariant tensor of $Sp(1)$, 
$$
\varepsilon^{\alpha_1\alpha_2} \varepsilon_{\beta_1\beta_2}
= \delta^{\alpha_1}_{[\beta_1} \delta^{\alpha_2}_{\beta_2]},
$$
we can see that
$$
\varepsilon^{a_1 a_2 \cdots a_{N_Q-3}} 
\bar{q}'{}_{a_1}{}^{\alpha_1 (I} \bar{q}'{}_{a_2}{}^{|\alpha_2| J)}
= {1\over2} \varepsilon^{a_1 a_2 \cdots a_{N_Q-3}} 
\bar{q}'{}_{a_1}{}^{\beta_1 (I} \bar{q}'{}_{a_2}{}^{|\beta_2| J)}
\varepsilon_{\beta_1\beta_2} \varepsilon^{\alpha_1\alpha_2},
$$
and it gives rise to the fifth.
This is always possible
when more than two $\bar{q}'$ from the second, the third, and the fourth
are contracted with the invariant tensor 
$\varepsilon^{a_1 \cdots a_{N_Q-3}}$ of $SU(N_Q-3)$,
because the global $SU(2)$ indices of two $\bar{q}'$ of them
must take the same value, thus symmetric.
Thus, we find that the total number of the second, the third, and the fourth
contracted with the same invariant tensor 
$\varepsilon^{a_1 \cdots a_{N_Q-3}}$ should be less than three.

When four of $\bar{q}'$ are contracted with the invariant tensor, 
each two of them take the same value of the $SU(2)$ indices, respectively, 
and can be rewritten in terms of two copies of the fifth and some other parts.
Thus, when one of the fifth is contracted with the invariant tensor, 
the total number of the second, the third, and the fourth
contracted with the same invariant tensor 
$\varepsilon^{a_1 \cdots a_{N_Q-3}}$ should be less than two.

Wrapping up these facts, 
together with the $F$-term conditions, \ConIII, and \ee,
we can verify that all the operators with the single 
$\varepsilon^{a_1 \cdots a_{N_Q-3}}$ are given by
\eqn\magopIV{
\eqalign{
&(*C)_{i_1 \cdots i_{N_Q-3}} \sim \varepsilon^{a_1 \cdots a_{N_Q-3}}
\bar{q}_{a_1i_1} \cdots \bar{q}_{a_{N_Q-3}i_{N_Q-3}}, 
\cr
&(*B)_{X i_1 \cdots i_{N_Q-5}} \sim
\varepsilon^{a_1 \cdots a_{N_Q-3}} \varepsilon^{\alpha\beta} \bar{q}_{a_1i_1} 
\cdots \bar{q}_{a_{N_Q-5}i_{N_Q-5}} \bar{q}'{}_{a_{N_Q-4}}^{\alpha I} 
(\sigma_2 \sigma_X)_{IJ} \bar{q}'{}_{a_{N_Q-3}}^{\beta J}, 
\cr
&(*F)_{i_1 \cdots i_{N_Q-7}} \sim
\varepsilon^{a_1 \cdots a_{N_Q-3}} \varepsilon_{\alpha\beta} 
\varepsilon_{\gamma\delta} \bar{q}_{a_1i_1} \cdots 
\bar{q}_{a_{N_Q-7}i_{N_Q-7}}, 
\cr
&\hskip 3cm \times \bar{q}'{}_{a_{N_Q-6}}^{\alpha I} (\sigma_2 \sigma_X)_{IJ}	\bar{q}'{}_{a_{N_Q-5}}^{\beta J} \bar{q}'{}_{a_{N_Q-4}}^{\gamma K} 
(\sigma_2 \sigma_X)_{KL} \bar{q}'{}_{a_{N_Q-3}}^{\delta L}, 
\cr
&(*H)_{i_1 \cdots i_{N_Q-4}} \sim
\varepsilon_{IJ} \varepsilon^{a_1 \cdots a_{N_Q-3}} \varepsilon_{\alpha\beta}
\bar{q}_{a_1i_1} \cdots \bar{q}_{a_{N_Q-4}i_{N_Q-4}} 
\bar{q}'{}_{a_{N_Q-3}}^{\alpha I} t^{\beta J}.
}}

We go on to the operators with more than two 
$\varepsilon^{a_1 \cdots a_{N_Q-3}}$ and skip those with two here.
The latter will be explained later.
We will see that all the operators in these classes 
do not give the independent gauge invariant operators.
Since the only fourth operators in \combinationII\ can connect with 
two invariant tensors $\varepsilon^{a_1 \cdots a_{N_Q-3}}$,
the operators with more than two $\varepsilon^{a_1 \cdots a_{N_Q-3}}$
should include at least one 
$\varepsilon^{a_1 \cdots a_{N_Q-3}}$ which 
are contracted with two of the fourth.
Further, all the remaining indices of the same 
$\varepsilon^{a_1 \cdots a_{N_Q-3}}$ must be contracted with the first
operator in \combinationII, as
\eqn\eqqq{
\eqalign{
\varepsilon^{a_1 \cdots a_{N_Q-3}} \bar{q}_{a_1} \cdots \bar{q}_{a_{N_Q-5}}
\left( \varepsilon_{\alpha\beta} \bar{q}'{}_{a_{N_Q-4}}{}^{\alpha (I}
\bar{q}'{}_{b_1}{}^{|\beta| J)} \varepsilon^{b_1 \cdots b_{N_Q-3}} \right) 
\cr
\times \left( \varepsilon_{\gamma\delta} \bar{q}'{}_{a_{N_Q-3}}{}^{\gamma (K}
\bar{q}'{}_{c_1}{}^{|\delta| L)} \varepsilon^{c_1 \cdots c_{N_Q-3}}
\right) ,
}}
as we can see from the previous discussion.
Here, we apply the identity
\eqn\eeII{
\varepsilon^{[a_1 \cdots a_{N_Q-3}} 
\varepsilon^{b_1] b_2 \cdots b_{N_Q-3}} = 0,
}
to $\varepsilon^{a_1 \cdots a_{N_Q-3}}$ and 
$\varepsilon^{b_1 \cdots b_{N_Q-3}}$ in \eqqq.
Ignoring the terms decomposed into the products of gauge invariant operators,
we find that the resulting operators are given by 
\eqn\eqq{
\eqalign{
{1\over2} \sum_{k=1}^{N_Q-5}
& \varepsilon^{a_1 \cdots a_{k-1} b_1 a_{k+1} \cdots a_{N_Q-3}}
\bar{q}_{a_1} \cdots \bar{q}_{a_{k-1}}
\bar{q}_{a_{k+1}} \cdots \bar{q}_{a_{N_Q-5}}
\left( \varepsilon_{\alpha\beta} \bar{q}'{}_{a_{N_Q-4}}{}^{\alpha (I}
\bar{q}'{}_{b_1}{}^{|\beta| J)} \right)
\cr
& \left( \varepsilon_{\gamma\delta} \bar{q}'{}_{a_{N_Q-3}}{}^{\gamma (K} 
\bar{q}'{}_{c_1}{}^{|\delta| L)} \varepsilon^{c_1 \cdots c_{N_Q-3}} \right)
\times \left( \varepsilon^{a_k b_2 \cdots b_{N_Q-3}} \bar{q}_{a_k} \right) .
}}
If the resulting operator is not decomposed into gauge invariant operators, 
the last factor $\varepsilon^{a_k b_2 \cdots b_{N_Q-3}} \bar{q}_{a_k}$ 
in \eqq\ are connected with the invariant tensor 
$\varepsilon^{c_1 \cdots c_{N_Q-3}}$ via other operators.
This happens only when the invariant tensor 
$\varepsilon^{c_1 \cdots c_{N_Q-3}}$ in \eqq\ is contracted with 
two of the fourth in \combinationII.
This is the same situation we previously have seen for the invariant tensor 
$\varepsilon^{a_1 \cdots a_{N_Q-3}}$ in \eqqq,
and thus, we can repeat the same procedure to show that the resulting 
operator is decomposed into gauge invariant operators.

We now turn to the operators with two $\varepsilon^{a_1 \cdots a_{N_Q-3}}$.
As discussed previously, the only fourth operator in \combinationII\
can be used to connect the two invariant tensors.
In particular, they are connected by at most two of the operator.
By using the identity \eeII, we can see that the invariant tensors
connected by two of the fourth in \combinationII\ can be reduced to 
those by one.
Thus, we only have to consider the latter operators.
If either of the invariant tensors does not have the fifth operator
of \combinationII, we can use the identity \eeII\ 
to show that they are decomposed into gauge invariant operators.
If both of them have the fifth operators, a closer examination 
is needed on the symmetry of the global $SU(2)$ indices of $\bar{q}'$s.
Taking account of this point and the identity \eeII, we can verify that 
they are also decomposed into gauge invariant operators.

To summarize, the singlets $M$, $Y$, and 
the operators listed in \magopI, \magopII, \magopIII, and \magopIV\ are
the gauge invariant generators 
of the classical chiral ring of the magnetic theory.

As discussed in section 2,
to all the gauge invariant generators \electricoperator\
of the classical chiral ring in the electric theory,
there exist the counterparts \matching\ in the magnetic theory. 
However, the extra gauge invariant operators \magop\ seem to exist
in the magnetic theory.
If the electric-magnetic duality is true for this model,
this discrepancy should disappear at the quantum level.


\listrefs

\end